\documentclass[12pt]{article}
\usepackage{graphicx}
\usepackage{enumerate}
\usepackage{comment}
\usepackage{setspace}
\usepackage{amsmath}
\usepackage{amssymb}
\usepackage[textwidth=16cm,textheight=22cm]{geometry}
\usepackage{color}

\begin{document}
\date{\today}
\begin{center}
  {\bf \Large Pion $p_T$ spectra in $p+p$ collisions as a function of $\sqrt{s}$ and event multiplicity}\\
\ \\
    {Priyanka Sett\footnote{email : sett.priyanka@gmail.com} and  Prashant Shukla\footnote{email : pshuklabarc@gmail.com}}\\
    {\it Nuclear Physics Division, Bhabha Atomic Research Center~$^{1, 2}$,\\
      \it Trombay, Mumbai - 400 085, India\\
      \it Homi Bhabha National Institute~$^{1, 2}$, Anushaktinagar, Mumbai - 400094, India\\}

\ \\
\ \\

{\bf Abstract}\\
\end{center}

  We study the charged pion transverse momentum ($p_T$) spectra
in $p$+$p$ collisions as a function of collision energy $\sqrt{s}$ and event 
multiplicity using Tsallis distribution. This study  
gives an insight of the pion production process in $p+p$ collisions. 
  The study covers pion spectra measured in $p+p$ collisions at SPS energies (6.27-17.27 GeV), 
RHIC energies (62.4 GeV and 200 GeV) and LHC energies (900 GeV, 2.76 TeV and 7 TeV).
  The Tsallis parameters have been obtained and parameterized 
as a function of $\sqrt{s}$. The study suggests that as we move to higher energy more
and more hard processes contribute to the spectra.
  We also study the charged pion spectra for different event multiplicities  
in $p$+$p$ collisions for LHC energies using Tsallis distribution.
  The variation of the Tsallis parameters as a function of event multiplicity has been 
obtained and their behavior is found to be independent of collision energy.

\section{Introduction}
\label{intro}

\indent

  The particle spectra measured in hadronic collisions are of utmost interest because of 
their fundamental nature and simplicity, which allow to verify pQCD~\cite{pQCD1, pQCD2} calculations 
and also help to make comprehensive phenomenological studies. 
  The ratios of the particle yields obtained from the measured spectra allow to get 
the chemical freeze-out conditions, whereas the spectra themselves reflect the conditions at 
the kinetic freeze-out. 
  The particle spectra provide useful information about the collision dynamics.
The low $p_T$ region of the spectrum corresponds to the particles originating from 
low momentum transfer and multi-scattering processes   
(non-perturbative QCD), whereas, the high $p_T$ region comes from the 
hard-parton-scattering (pQCD) among the initial partons. The transition of this non-perturbative 
to perturbative dynamics has no sharp boundary, though one can have an estimate from the 
'$x_T-scaling$'~\cite{x_T_scaling}.  
  Extensive~\cite{extensive1, extensive2} and 
non-extensive~\cite{Tsallis, non-extensive1, non-extensive2, non-extensive3, urmossy} 
statistical approaches have been used to characterize particle spectra in terms 
of thermodynamic variables.  
  Extensive statistics assume thermal and chemical equilibrium of the system at hadronic phase which  
lead to an exponential distribution of the particle spectra. 
  In experiments, the particle spectra show a power-law behavior at high $p_T$. 
This behavior is reproduced by the non-extensive approach with an additional parameter.
  In recent times, the Tsallis~\cite{Tsallis} statistical approach 
is widely used to describe the particle spectra obtained in high-energy collisions
with only two parameters; the temperature $T$ and $q$, known as non-extensivity parameter which is 
a measure of temperature fluctuations or degree of non-equilibrium in the system.  

  The Tsallis distribution gives an excellent description of $p_{T}$ spectra 
of all identified mesons measured in $p+p$ collisions at $\sqrt{s}=200$\,GeV~\cite{PPG099}.
In a recent work~\cite{khandai_ijmpa, tsallis_sps}, the Tsallis distribution has been used to describe the 
$p_T$ spectra of identified charged hadrons measured in $p+p$ collisions at RHIC and at LHC energies. 
Such an approach has also been applied to the inclusive charged hadron $p+p$ data in  
recent publications~\cite{tsallis_ch_hadrons1, tsallis_ch_hadrons2}. 
   It has been shown in Ref.~\cite{khandai_ijmpa, wong_wilk} that the 
functional form of the Tsallis distribution with thermodynamic origin is of the same form as 
the QCD-inspired Hagedorn formula~\cite{Hagedorn, HAGEFACT}. This could be the reason of success of 
Tsallis distribution in $p+p$ collisions which is a power law typical of QCD hard scatterings.
 The hardness of the spectra is thus related to $q$ and the parameter $T$ governs the contribution
from soft collisions. 

  Using the Tsallis phenomenological function, we review and study the charged pion 
spectra in $p+p$ collisions in a large energy regime, spanning from  
SPS ~\cite{na61} (6.27 GeV - 17.27 GeV), RHIC~\cite{PPG101} (62.4 and 200 GeV) 
to LHC~\cite{cmspp_900_2.76_7} (900 GeV, 2.76 TeV and 7 TeV) energies.  
  The object of the present work is to study the behaviour of the Tsallis parameters as a function
of collision energy. We also study the charged pion spectra for different event multiplicities  
in $p$+$p$ collisions for LHC energies.
Among all hadrons, pions are chosen because of their abundance in collisions, 
simple quark structure and availability of the data at different energies.

\section{Formalism}
\label{formulation} 

\indent 

 The transverse momentum spectra of hadrons, obtained from different fixed and 
collider experiments have shown that, the high $p_T$ region of the spectra 
can be described successfully by the power law,
\begin{eqnarray}
\label{eqpower}
E \frac{d^3N}{dp^3} =  C_P p_T^{-n},
\end{eqnarray}
where $C_P$ is the normalization constant and $n$ is the power which determines the shape of the spectra 
at high $p_T$. 
However, the low $p_T$ region of the particle spectra shows an exponential shape   
and can be described by the Boltzmann-Gibbs~\cite {BG_statistics1, BG_statistics2} statistical approach, 
\begin{eqnarray}
\label{eqboltzman}
E \frac{d^3N}{dp^3} = C_B e^{-E/T},
\end{eqnarray}
where $C_B$ is the normalization constant, $E$ is the particle energy and $T$ is the temperature of the system. 

  In the early 80's, Hagedorn~\cite{Hagedorn} proposed a phenomenological function which 
describes the particle spectra for both the higher and lower $p_T$ regions:
\begin{eqnarray}
\label{eqhage}
E \frac{d^3N}{dp^3} = A \left(1 + \frac{p_T}{p_0}\right)^{-n},
\end{eqnarray}
where $A$, $p_0$ and $n$ are the fit parameters. 
The above equation describes an exponential behavior for low $p_T$ and a power-law 
behavior for high $p_T$. 
\begin{eqnarray}
  \left(1 + {p_{T}\over p_0}\right)^{-n}    & \simeq &  {\rm exp}\left(\frac {-np_{T}} {p_0}\right), \, \, \, \,\,  {\rm for}\,\,\, p_{T} \rightarrow 0 \\
  & \simeq  &\left(\frac {p_0} {p_{T}}\right)^n,  \,\,\,\,\,\,{\rm for}\,\,\, p_{T} \rightarrow \infty. 
\end{eqnarray}  
  The parameter $n$ in this equation is often related to the 'power' in the 
'QCD-inspired' quark interchange model~\cite{HAGEFACT}. 

  In the late 80's, Tsallis ~\cite{Tsallis} introduced the idea of the non-extensive statistics in place 
of thermal Boltzmann-Gibbs statistics. This approach includes a parameter $q$, known as non-extensive 
parameter which quantifies the temperature fluctuation~\cite{q_Tsallis} in the system as : 
$q-1 = Var(1/T)/\langle T \rangle^{2}$. 
 The non-extensive statistics assume Boltzmann-Gibbs form in the limit $q$ $\rightarrow$ 1. 
In Tsallis approach, the Boltzmann-Gibbs distribution takes the form 
\begin{eqnarray}
 E{d^3N \over dp^3}  =  C_{q}\left(1 + (q-1)\frac {E}{T}\right)^\frac{-1}{q-1},
\label{Tsallis_q}
\end{eqnarray}
where $C_q$ is the normalization factor.
  One can use the relation $E$ = $m_T$ at mid-rapidity and $n$ = $1/(q-1)$ in Eq.~\ref{Tsallis_q} to obtain :
\begin{eqnarray}
 E{d^3N \over dp^3}  =  C_{n}\left(1 + \frac {m_{T}}{nT}\right)^{-n},
\label{Tsallis1}
\end{eqnarray}
where, $C_n$ is the normalization factor.  
Eq.~\ref{Tsallis1} can be re-written as : 
\begin{eqnarray}
   \frac{1}{2\pi p_T} \frac{d^{2}N}{dp_T dy} = C_{n}\left(1 + \frac {m_{T}}{nT}\right)^{-n}, 
\label{deriv1}
\end{eqnarray}
The value of $C_n$ can be obtained by integrating Eq.~\ref{deriv1} over momentum space :
 \begin{eqnarray}
   C_n = \frac{dN/dy}{\int^{\infty}_{0} \left(1 + \frac {m_{T}}{nT}\right)^{-n}  2 \pi p_T dp_T}, 
\label{deriv2}
\end{eqnarray}
Here the quantity $dN/dy$ is the $p_T$ integrated yield.  
Eq.~\ref{Tsallis1} with the normalization constant takes the form~\cite{PPG099} : 
\begin{eqnarray}
  E\frac{d^3N}{dp^3}  & = & \frac{1}{2\pi} \frac{dN}{dy} \frac{(n-1)(n-2)}{(nT+m(n-1))(nT+m)} \left(\frac{nT+m_T}{nT+m}\right)^{-n}, 
\label{Tsallis}
\end{eqnarray}

 Larger values of $q$ correspond to smaller values of $n$ which imply dominant hard 
QCD point-like scattering. Both $n$ and $q$ have been interchangeably used in 
Tsallis distribution~\cite{non-extensive1, PPG099, PPG101, Star, cleymans}. 
The Tsallis interpretation of parameters $T$ as temperature and $q$ as non-extensivity parameter
is more suited for heavy ion collisions while for $p+p$ collisions Hagedorn interpretation in terms of 
power $n$ and a parameter $T=p_0/n$ which controls soft physics processes is more meaningful. 
 Phenomenological studies suggest that, for quark-quark point scattering, $n\sim$ 4~\cite{BlankenbeclerPRD12, BrodskyPLB637}  
and when multiple scattering centers are involved $n$ grows larger.

There are many other forms of Eq.~\ref{Tsallis}, which are used often to describe particle spectra 
in literature, see e.g. Refs.~\cite{non-extensive1, PPG101, Star, cleymans, Tsallis-Pareto, Tsallis-Levy,  ppbaryon}. 
% Refs.~\cite{non-extensive1, cleymans, Tsallis-Pareto, Tsallis-Levy, Star, ppbaryon, PPG101}. 

\section{Results and Discussions}

\indent  

    All the studies are performed with Eq.~\ref{Tsallis} and the fit parameters $n$, $T$ 
and $dN/dy$ are obtained. 
     The different experiments, energies, rapidity ranges and particles 
used in the analysis are summarized in Table~\ref{expts}. 
For SPS energies only available data is for $\pi^{-}$ measured by NA61 Collaboration~\cite{na61}, 
for RHIC and LHC energies we use ($\pi^{+}$ + $\pi^{-}$)/2. 
  All the data used are measured in mid-rapidity and are given for unit rapidity.
The difference in rapidity range is not expected to affect the behaviour of the spectra.
 CMS experiment presented~\cite{cmspp_900_2.76_7} transverse momentum spectra for different events classified 
on the basis of number of true tracks referred here as track multiplicity of event or simply 
multiplicity. Each multiplicity class is represented by average number of tracks
($\langle N_{tracks} \rangle$).

\begin{table}
\begin{center}
\caption{The center of mass energy and rapidity of the data used for the study.}
\label{expts}
\begin{tabular}{cccc}
\hline
Experiments &  Center of mass                     & Rapidity & Particles  \\
            &   energy (GeV)                      &          & Studied   \\  
\hline
\hline
SPS         &   6.27, 7.74,                       & 0.0--0.2 & $\pi^{-}$ \\
            &   8.76, 12.32, 17.27                &          &           \\ 
            &                                     &          &            \\   

%\hline
RHIC        &  62.4, 200                         & $|y| <$ 0.35 & $\pi^{+}$, $\pi^{-}$ \\
%\hline
            &                                    &          &            \\   
LHC         &  900, 2760, 700                    & $|y| <$ 1.0  & $\pi^{+}$, $\pi^{-}$ \\
\hline
\end{tabular}
\end{center}
\end{table}

% sps_rhic_lhc_pion plots for pions in pp :
\begin{figure}
\begin{center}
\includegraphics[width=0.48\textwidth]{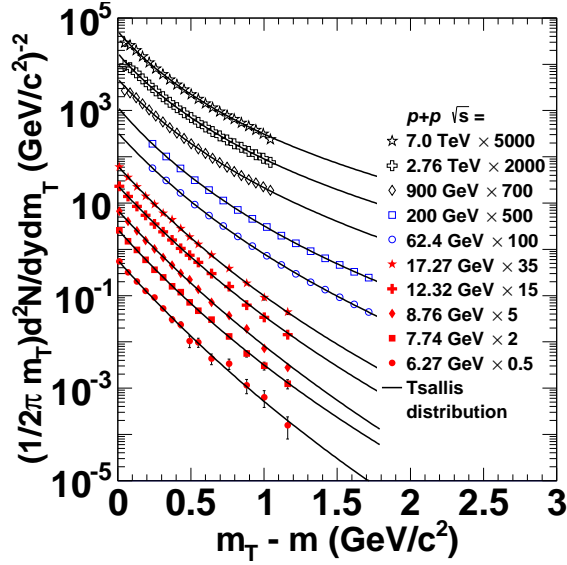}
\caption{(Color online) The invariant yield spectra of charged pions as a function of ($m_T - m$) for SPS~\cite{na61} energies 6.27 GeV, 
7.74 GeV, 8.76 GeV, 12.32 GeV and 17.27 GeV, RHIC~\cite{PPG101} energies 62.4 GeV and 200 GeV and LHC~\cite{cmspp_900_2.76_7} 
energies 900 GeV, 2.76 TeV and 7 TeV. The solid lines are the Tsallis function (Eq.~\ref{Tsallis}). The negative pion yields are 
plotted for SPS energies and for all other energies, average yield for positive and negative pion are plotted.}
\label{fig:pt_pp_sps_rhic_lhc_pion} 
\end{center} 
\end{figure}

\begin{figure}
\begin{center}
\includegraphics[width=0.48\textwidth]{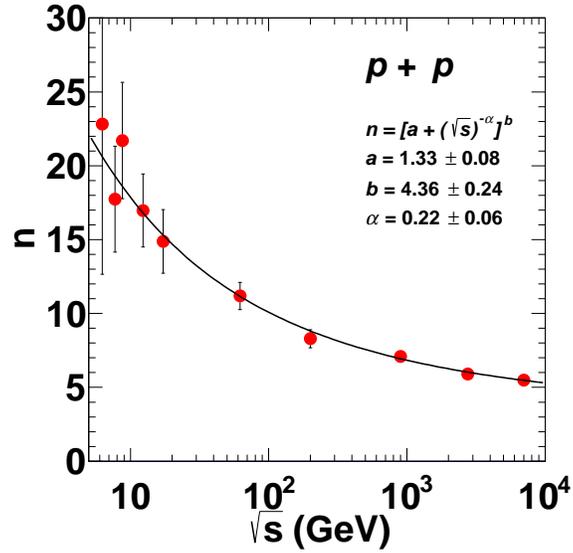}
\caption{(Color online) The variation of the Tsallis parameter $n$ for charged pions as a function of $\sqrt{s}$. The solid curve 
  represents the parameterization $\left(a + (\sqrt{s})^{-\alpha}\right)^{b}$.}
\label{fig:n_pp_sps_rhic_lhc_pion} 
\end{center} 
\end{figure}

\subsection{Tsallis parameters as a function of $\sqrt{s}$ in $p+p$ system :}
\label{n_T_pp_roots}

\indent

  In this analysis all the Tsallis parameters  are obtained for 
charged pion spectra as a function of $\sqrt{s}$ in $p+p$ system for SPS~\cite{na61}, RHIC~\cite{PPG101} and 
LHC~\cite{cmspp_900_2.76_7} energies.
  Similar study is available in Ref.~\cite{khandai_ijmpa} using RHIC and LHC data and in 
Ref.~\cite{tsallis_sps} for SPS and LHC data. 

   The pion $p_T$ spectra measured in $p+p$ collisions at different $\sqrt{s}$ are shown in
Fig.~\ref{fig:pt_pp_sps_rhic_lhc_pion} along with with Tsallis fits (Eq.~\ref{Tsallis}) shown  
by solid lines. The spectra are scaled by arbitrary factors (given in figure) for visual clarity. 
In case of RHIC data, we restrict the $p_T$ range to 1.7 GeV/$c^2$ to have similar $p_T$ range at all energies.  
 It can be noticed that the spectra become harder with increase in $\sqrt{s}$ 
which is depictive of occurrence of harder scatterings at higher collision energy.
  The $\chi^{2}$ per degree of freedom $\chi^{2}/NDF$ values for all the fits are given in 
Table~\ref{chi2table1}. The $\chi^{2}/NDF$ values are $\lesssim$ 1, which is indicative of 
good fit quality.

\begin{table}
\begin{center}
\caption{Values of the $\chi^{2}/NDF$ for Tsallis fits of pion spectra at different $\sqrt{s}$. }
\label{chi2table1}
\begin{tabular}[width=0.45\textwidth]{cc}
\hline
$\sqrt{s}$  & $\chi^{2}/NDF$ \\
\hline
\hline
6.27 GeV & 6.31/12\\
7.74 GeV & 5.80/12\\
8.76 GeV & 10.68/12\\
12.32 GeV & 9.25/12\\
17.27 GeV & 2.65/12\\
62 GeV & 0.74/11\\
200 GeV & 0.48/11\\
900 GeV & 24.33/19\\
2.76 TeV & 5.59/19\\
7.00 TeV & 13.11/19\\
\hline
\end{tabular}
\end{center}
\end{table}
\begin{figure}
\begin{center}
\includegraphics[width=0.48\textwidth]{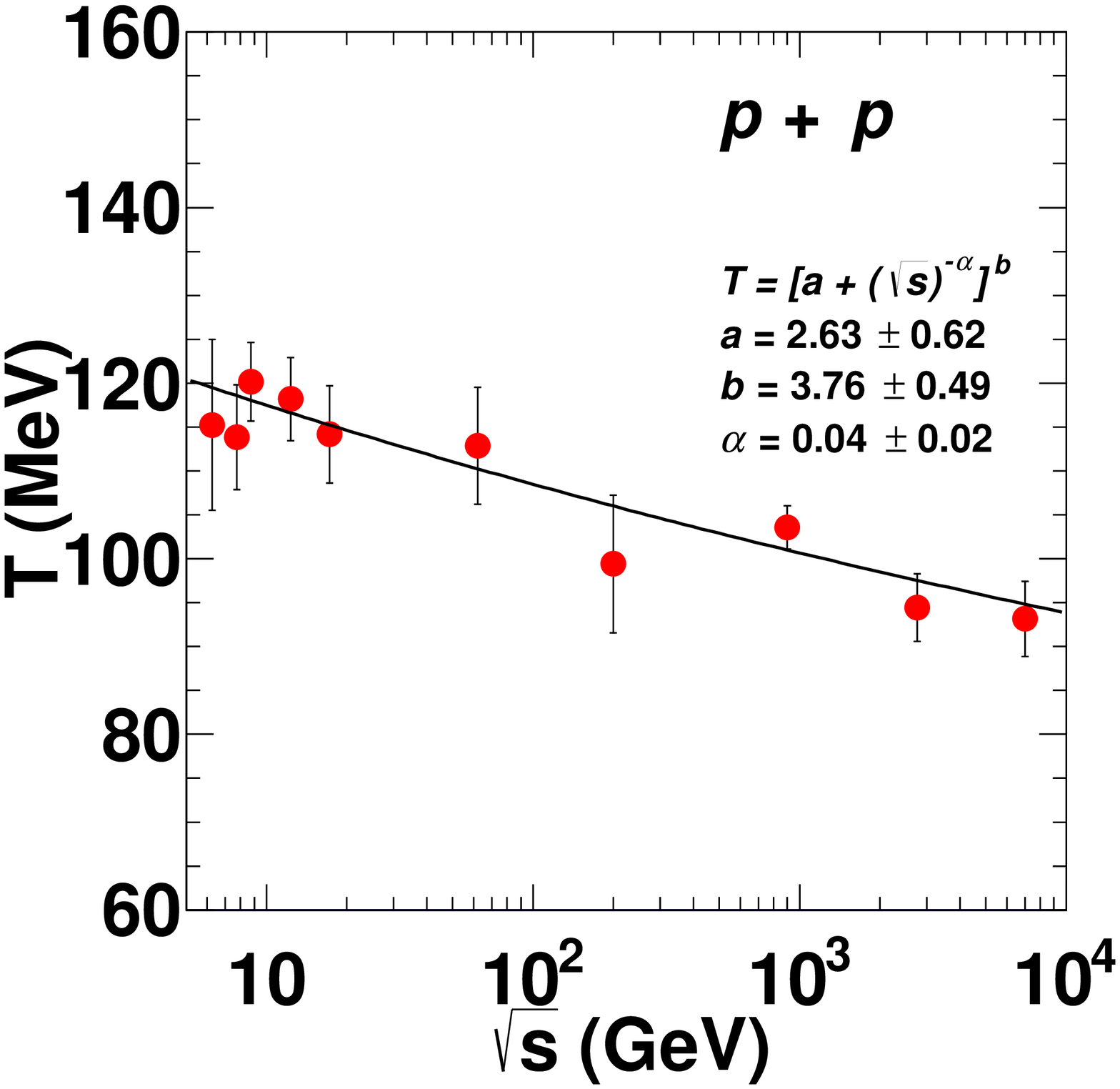}
\caption{(Color online) The variation of the Tsallis parameter $T$ for charged pions as a function of $\sqrt{s}$. The solid curve 
  represents the parameterization $\left(a + (\sqrt{s})^{-\alpha}\right)^{b}$.}
\label{fig:T_pp_sps_rhic_lhc_pion} 
\end{center} 
\end{figure}

\begin{figure}
\begin{center}
\includegraphics[width=0.48\textwidth]{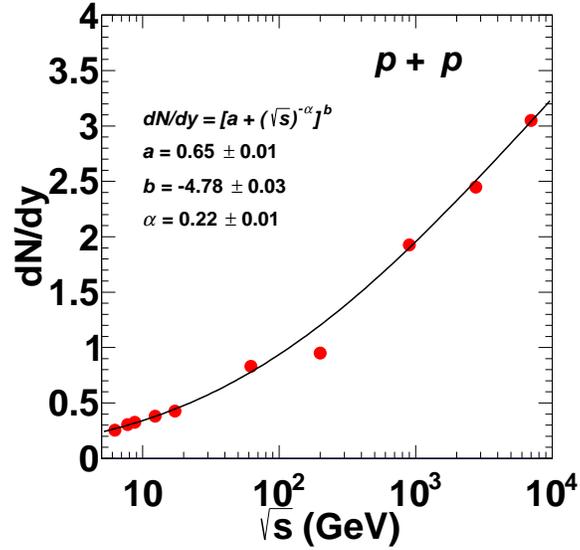}
\caption{(Color online) The variation of the integrated yield $dN/dy$ for charged pions as a function of $\sqrt{s}$. The solid curve 
  represents the parameterization $\left(a + (\sqrt{s})^{-\alpha}\right)^{b}$.}
\label{fig:dndy_pp_sps_rhic_lhc_pion} 
\end{center} 
\end{figure}

   The parameters $n$ and $T$ obtained from this analysis are shown in 
Fig.~\ref{fig:n_pp_sps_rhic_lhc_pion} and Fig.~\ref{fig:T_pp_sps_rhic_lhc_pion}, 
respectively,  as a function of $\sqrt{s}$. 
 The variation of $dN/dy$ as a function of $\sqrt{s}$ is shown in Fig.~\ref{fig:dndy_pp_sps_rhic_lhc_pion}. 
 The parameter $n$ decreases with increasing $\sqrt{s}$ and starts saturating at LHC energies. 
The value of $T$ also reduces slowly from SPS energies to LHC energies. 
 The integrated yield $dN/dy$ increases 10 times when going from SPS to highest LHC energy.
 
  Larger value of $n$ (also larger value of $T$) suggests that the spectra has 
contribution from processes involving small momentum transfer  
arising due to the re-scattering, recombination of partons, fragmentation from strings etc.
 Whereas, smaller values of $n$ are indicative of harder processes are involved in 
particle-production. Thus the spectra at SPS energies have large softer contribution and as
the collision energy increases more and more contribution from hard processes are added.
  
  All the three parameters can be parametrized by a function of type 
\begin{eqnarray}
\label{par0}
f(\sqrt{s}) = \left(a + (\sqrt{s})^{-\alpha}\right)^{b}
\end{eqnarray}
Here $a=1.33 \pm 0.08$,  $\alpha =  0.22 \pm 0.06$ and $b=4.36 \pm 0.24$ for $n(\sqrt{s})$, 
$a=2.63 \pm 0.62$,  $\alpha =  0.04 \pm 0.02$ and $b=3.76 \pm 0.49$ for $T(\sqrt{s})$ and 
$a=0.65 \pm 0.01$,  $\alpha =  0.22 \pm 0.01$ and $b=-4.78 \pm 0.03$ for $dN(\sqrt{s})/dy$. 
Using the parameterizations for $n$ by Eq.~\ref{par0} we get  $n$ $\sim$ 3.46 
in the limit $\sqrt{s} \rightarrow \infty$.  
  The extrapolated values for $n$, $T$ and $dN/dy$ for $\sqrt{s}$ = 14 TeV are, 
$n$ $\sim$ 5.09, $T$ $\sim$ 90.33 MeV and $dN/dy$ $\sim$ 3.44.

\subsection{Tsallis parameters as a function of multiplicity ($\langle N_{tracks} \rangle$) for LHC energies :} 
\label{n_T_pp_multi_roots}
 
\indent 

 The Tsallis parameters for charged pion spectra are studied as a function of event multiplicity  
for different LHC energies 900 GeV, 2.76 and 7 TeV. The event multiplicity data was also studied 
in a recent work~\cite{urmossy} but our analysis and interpretations are different.
 
\begin{figure}
\begin{center}
\includegraphics[width=0.48\textwidth]{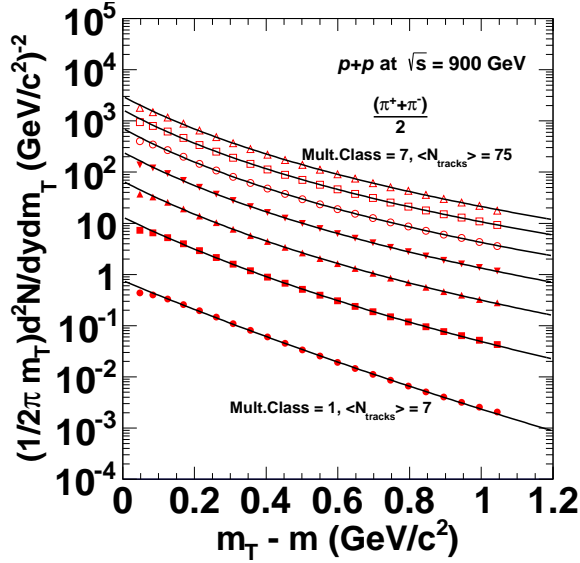}
\caption{(Color online) The invariant yield spectra of ($\pi^{+}+\pi^{-}$)/2~\cite{cmspp_900_2.76_7}, as a function of $m_T - m$ for $p+p$ 
  collisions at $\sqrt{s}$ = 900 GeV. The yields are shown for $\langle N_{tracks} \rangle$ 7, 16, 28, 40, 52, 63 and 75. 
  The spectra are scaled up for clarity by a factor of 6$^{i}$, where i = 0, 1, 2, 
  3, 4, 5 and 6. The solid lines show the Tsallis fits.}
\label{fig:pp_pion_mult_900gev} 
\end{center} 
\end{figure}

\begin{figure}
\begin{center}
\includegraphics[width=0.48\textwidth]{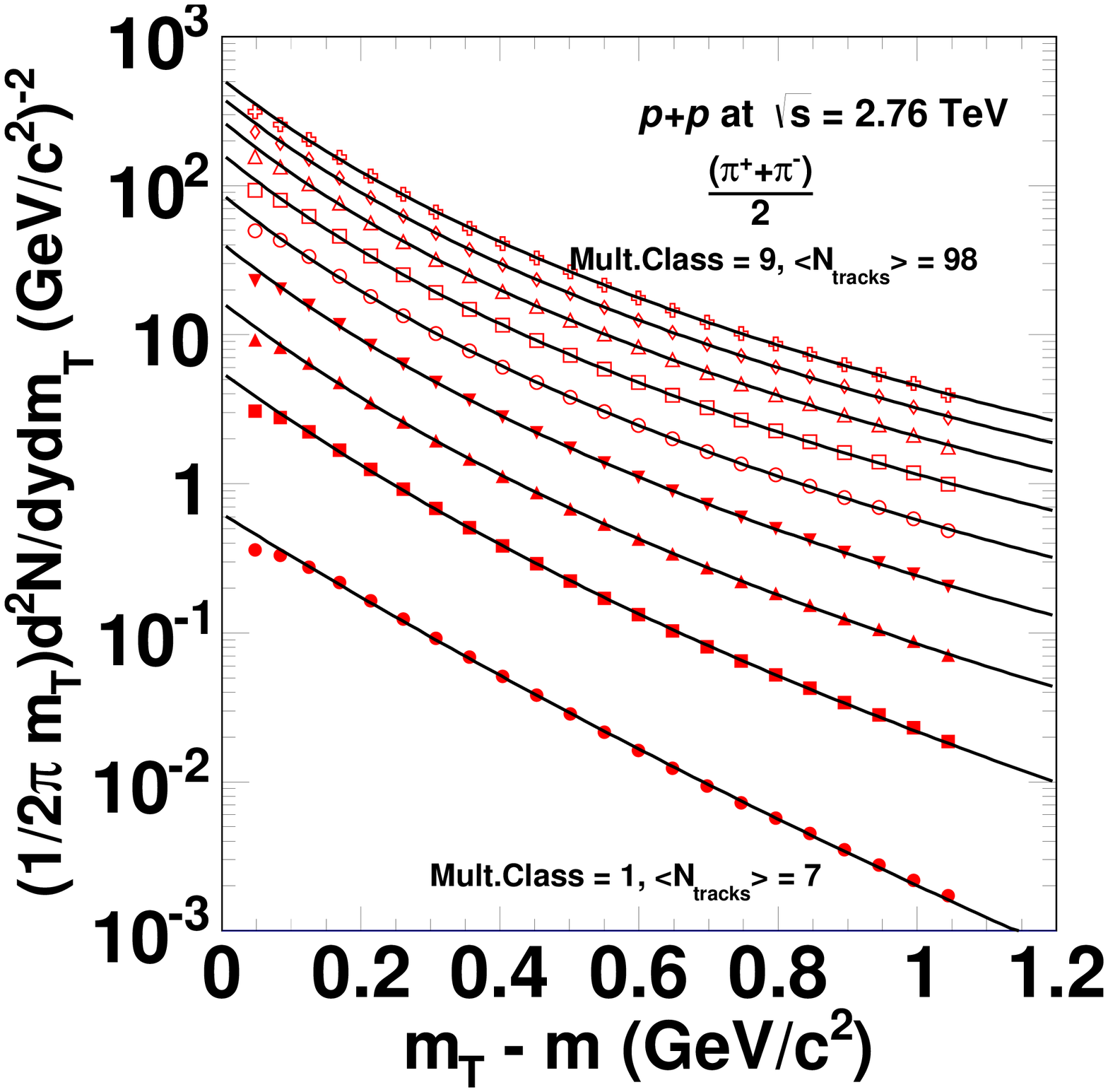}
\caption{(Color online) The invariant yield spectra of ($\pi^{+}+\pi^{-}$)/2~\cite{cmspp_900_2.76_7}, as a function of $m_T - m$ for $p+p$ 
  collisions at $\sqrt{s}$ = 2.76 TeV. The yields are shown for $\langle N_{tracks} \rangle$ 7, 16, 28, 40, 52, 63, 75, 86 and 98. 
  The spectra are scaled up for clarity by a factor of 3$^{i}$, where i = 0, 1, 2, 3, 4, 5, 6, 7 and 8. 
  The solid lines show the Tsallis fits.}
\label{fig:pp_pion_mult_2760gev} 
\end{center} 
\end{figure}

\begin{figure}
\begin{center}
\includegraphics[width=0.48\textwidth]{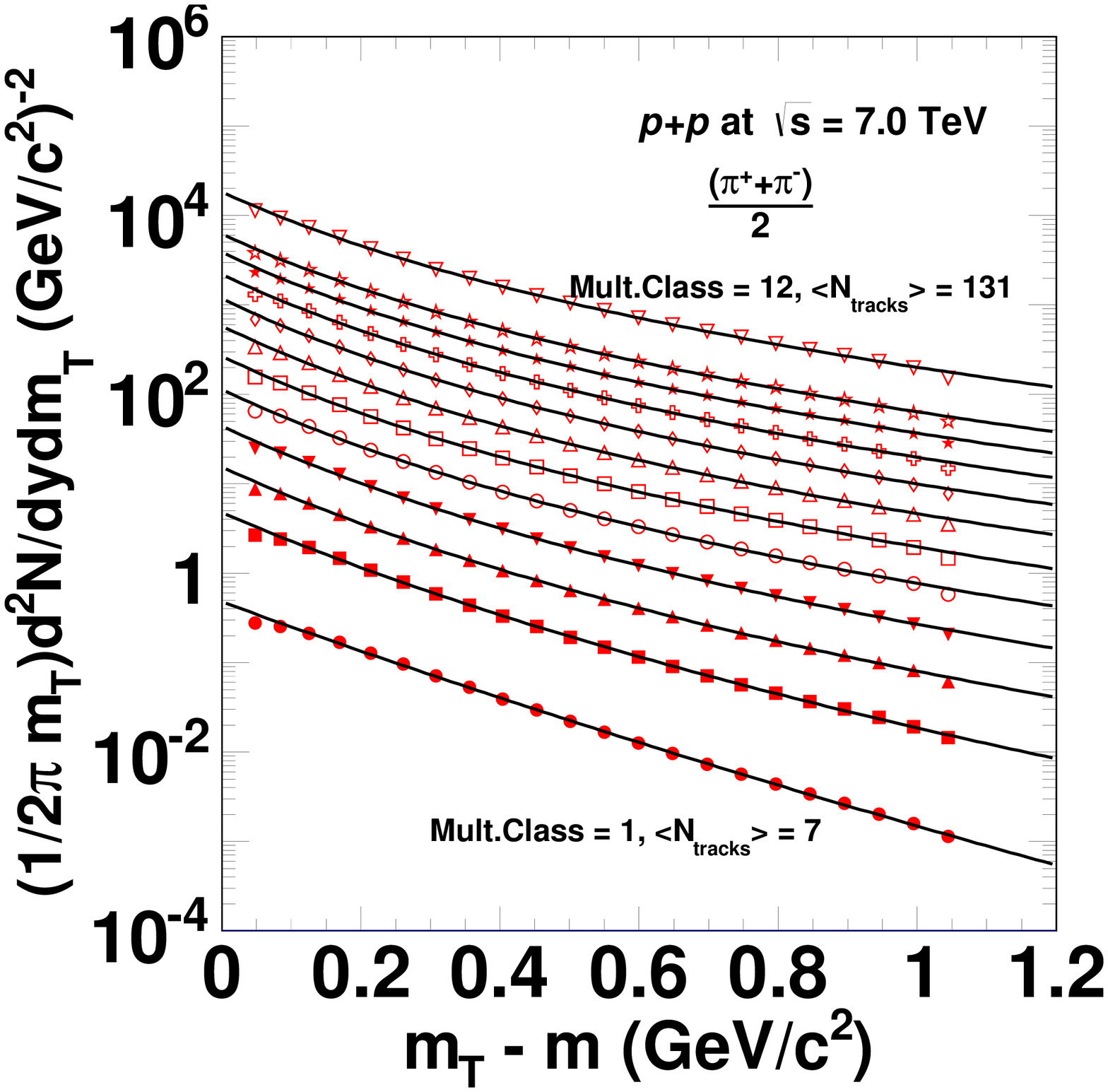}
\caption{(Color online) The invariant yield spectra of ($\pi^{+}+\pi^{-}$)/2~\cite{cmspp_900_2.76_7}, as a function of $m_T - m$ for $p+p$ 
  collisions at $\sqrt{s}$ = 7.00 TeV. The yields are shown for $\langle N_{tracks} \rangle$ 7, 16, 28, 40, 52, 63, 75, 
  86, 98, 109, 120 and 131. The spectra are scaled up for clarity by a factor of 3$^{i}$, where i = 
  0, 1, 2, 3, 4, 5, 6, 7, 8, 9, 10 and 11.   The solid lines show the Tsallis fits.}
\label{fig:pp_pion_mult_7000gev} 
\end{center} 
\end{figure}

 The invariant yield spectra corresponding to different multiplicities are fitted with Eq.~\ref{Tsallis}, 
are shown by the solid black lines in Fig.~\ref{fig:pp_pion_mult_900gev} for 900 GeV, in Fig.~\ref{fig:pp_pion_mult_2760gev} for 
2.76 TeV and in Fig.~\ref{fig:pp_pion_mult_7000gev} for 7 TeV center of mass energy.
The spectra are scaled up for distinctness. 
  The Tsallis distribution describes all the spectra well,  shown by
the $\chi^{2}/NDF$ values given in Table~\ref{chi2table2}.
The $\chi^{2}/NDF$ values are little higher for some of the lower  multiplicities due to the deviation of first data point 
in $p_T$  spectra with the curve.

% cms plots for pp  :----------------- only cms results as a func of multiplicty. ----------------------------

\begin{table}
\begin{center}
\caption{Values of the $\chi^{2}/NDF$ for the Tsallis fits in different event multiplicities.}
\label{chi2table2}
\begin{tabular}[width=0.45\textwidth]{cccc}
\hline
$\langle N_{tracks} \rangle$ &  \multicolumn{3}{c}{$\chi^{2}/NDF$ values for} \\
\multicolumn{1}{c}{}      &  900 GeV & 2.76 TeV & 7.0 TeV  \\ 
\hline 
\hline
7 &  48.89/18 & 57.66/18  & 48.10/18  \\
16 & 25.21/18 & 26.38/18  & 27.76/18  \\
28 & 9.97/18  & 11.47/18  & 22.83/18 \\ 
40 & 7.20/18  & 6.91/18   & 19.25/18\\
52 & 8.00/18  & 6.08/18   & 22.52/18 \\
63 & 9.18/18  & 6.44/18   & 26.30/18  \\
75 & 15.01/18 & 9.05/18   & 21.30/18 \\
86 &          & 8.16/18   & 19.24/18  \\
98 &          & 11.91/18  & 23.59/18 \\
109 &         &           & 20.82/18 \\
120 &         &           & 16.85/18\\
131 &         &           & 19.77/18 \\
\hline
\end{tabular}
\end{center}
\end{table}

\begin{figure}
\begin{center}
\includegraphics[width=0.48\textwidth]{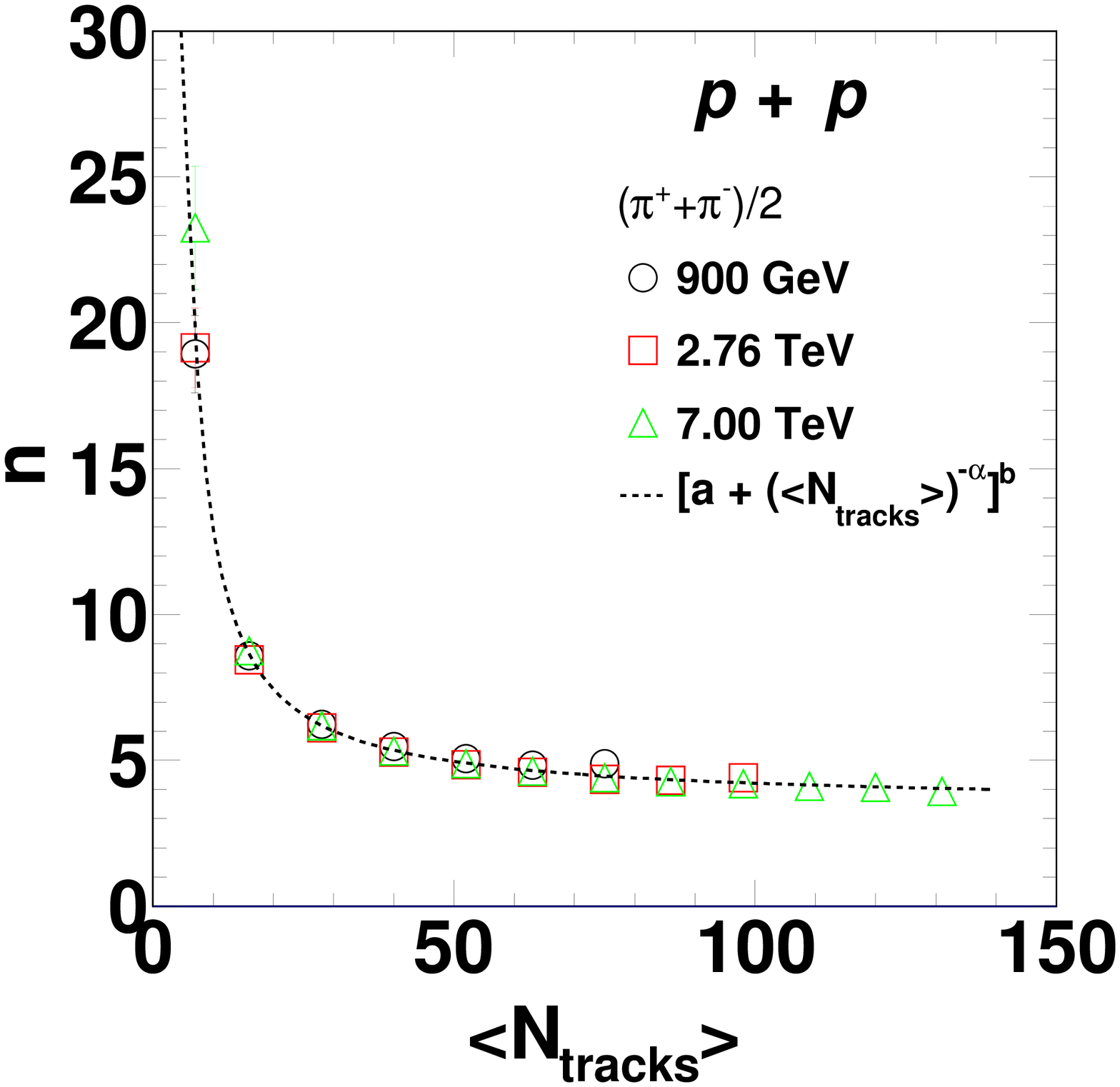}
\caption{(Color online) The variation of the Tsallis parameter $n$ for charged pions as a function of $\langle N_{tracks} \rangle$. 
  The variation is shown for 900 GeV by black circles, 2.76 TeV by red squares and 7.00 TeV by green triangles. The dashed curve 
  represents the parameterization $\left(a + (\langle N_{tracks} \rangle)^{-\alpha}\right)^{b}$.}
\label{fig:pion_n_ppmult} 
\end{center} 
\end{figure}

\begin{figure}
\begin{center}
\includegraphics[width=0.48\textwidth]{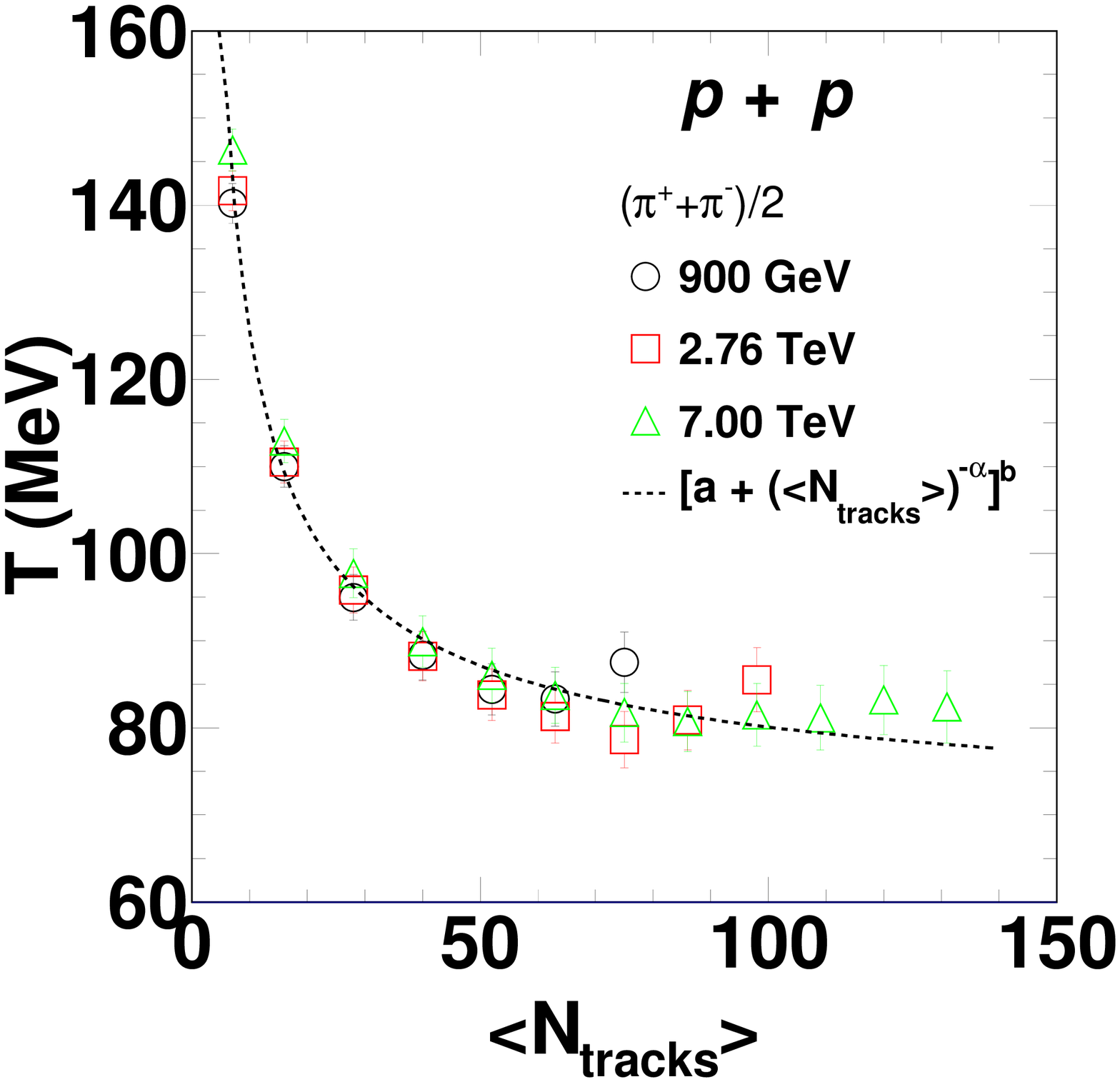}
\caption{(Color online) The variation of the Tsallis parameter $T$ for charged pions as a function of $\langle N_{tracks} \rangle$. 
  The variation is shown for 900 GeV by black circles, 2.76 TeV by red squares and 7.00 TeV by green triangles. The dashed curve 
  represents the parameterization $\left(a + (\langle N_{tracks} \rangle)^{-\alpha}\right)^{b}$.}
\label{fig:pion_T_ppmult} 
\end{center} 
\end{figure}

\begin{figure}
\begin{center}
\includegraphics[width=0.48\textwidth]{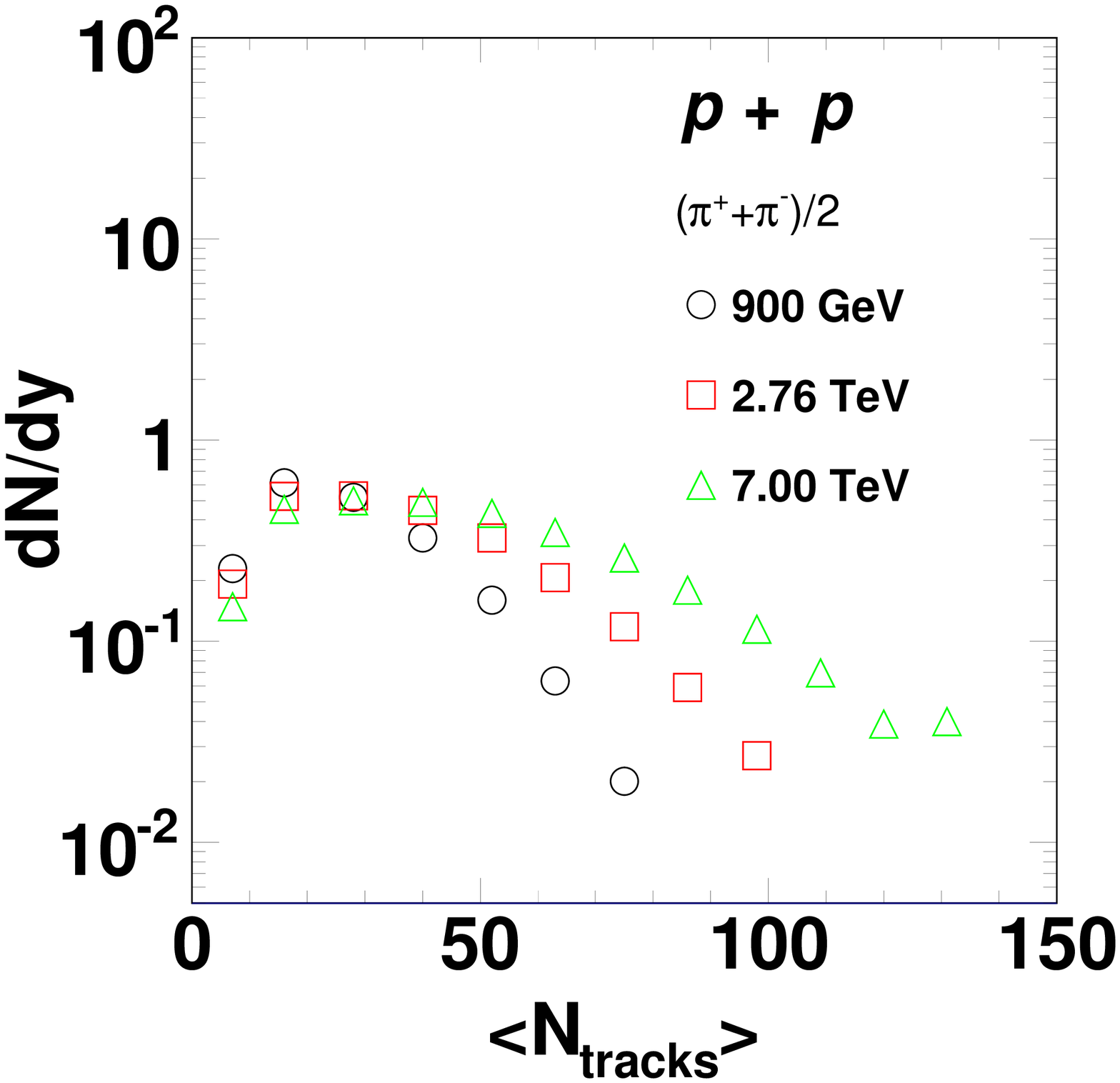}
\caption{(Color online) The variation of the integrated yield $dN/dy$ for charged pions as a function of $\langle N_{tracks} \rangle$. 
  The variation is shown for 900 GeV by black circles, 2.76 TeV by red squares and 7.00 TeV by green triangles.}
\label{fig:pion_dndy_ppmult} 
\end{center} 
\end{figure}

  The parameters $n$ and $T$  obtained from the fits are shown in Fig.~\ref{fig:pion_n_ppmult} and 
Fig.~\ref{fig:pion_T_ppmult} respectively, as a function of $\langle N_{tracks} \rangle$. 
The circles, squares and triangles correspond to the parameter values 
obtained from data at 900 GeV, 2.76 TeV and 7 TeV, respectively. 
  It is seen that both the parameters $n$ and $T$ decrease rapidly and then start saturating with the 
increase of $\langle N_{tracks} \rangle$ for all three energies. This variation (of $n$ and $T$) is very similar to 
the variation which we find as a function of $\sqrt{s}$.
  It means that events with higher multiplicity have larger contribution from hard processes.
 The value of $n$ for high multiplicity events at 7 TeV is $\sim$ 4 which is depictive of 
production from point quark-quark scattering.  
   The variation of $n$ and $T$ as a function of $\langle N_{tracks} \rangle$ can be described by the 
same curve given in the figure for all three energies and are 
parameterized by, 
\begin{eqnarray}
\label{par1}
f((<N_{tracks}>) = \left(a + (<N_{tracks}>)^{-\alpha}\right)^{b}
\end{eqnarray}
Here $a=1.13 \pm 0.01$,  $\alpha =  0.81 \pm 0.04$ and $b=10.32 \pm 0.76$ for $n(\langle N_{tracks} \rangle)$ and
$a=2.20 \pm 0.06$,  $\alpha =  0.56 \pm 0.08$ and $b=5.33 \pm 0.23$ for $T(\langle N_{tracks} \rangle)$.

  The $p_T$ integrated pion yield distribution in different multiplicity classes 
is shown in Fig.~\ref{fig:pion_dndy_ppmult} for the three LHC energies.
 The total $p_T$ integrated pion yield for each energy can be obtained by integrating the above
distributions over all multiplicity classes.
  It is noticed that as the energy increases more and more high mutliplicity events are added in 
the sample with mean of the distribution shifting towards higher $\langle N_{tracks} \rangle$.

\section{Conclusion}

\indent 

  This work presented the study of the transverse momentum spectra of the charged pions for different 
collisional energies and also for different event-multiplicities (at LHC energies) using Tsallis 
distribution.
 The Tsallis parameter $n$ decreases with increasing $\sqrt{s}$ and starts saturating at LHC energies. 
The value of $T$ also reduces slowly from SPS energies to LHC energies. 
It means that the spectra at SPS energies have large softer contribution and as
the collision energy increases more and more contribution from hard processes are added.
  The $p_T$ integrated pion yield increases with increasing $\sqrt{s}$ and 
becomes 10 times when going from SPS to highest LHC energy. 
   The Tsallis parameters are also  obtained as a function of event multiplicity
for all three LHC energies which can be described by the same curve.  
 The variation of $n$ and $T$ as a function of multiplicity is very similar to 
the variation which we find as a function of $\sqrt{s}$.
  It means that events with higher multiplicity have larger contribution from hard processes.
 The value of $n$ for high multiplicity events at 7 TeV is $\sim$ 4 which is depictive of 
production from point quark-quark scattering.  
   The $p_T$ integrated pion yield distribution for the three LHC energies shows that as the 
energy increases, more and more high mutliplicity events are added in the sample with mean of 
the distribution shifting towards higher multiplicity.

%\section{References}


\begin{thebibliography}{50} 
  

\bibitem{pQCD1} G. Sterman {\it et~al.}  (CTEQ Collaboration), Rev. Mod. Phys. {\bf 67}, 157 (1995).

\bibitem{pQCD2}  J.F. Owens {\it et~al.}, Phys. Rev. D{\bf 18}, 1501 (1978). 

\bibitem{x_T_scaling} R. Blankenbecler, S.J. Brodsky and J.F. Gunion, Phys. Lett. B{\bf 42}, 461 (1972).

\bibitem{extensive1} E. Fermi, Progress Theor. Phys., {\bf 5}, 570 (1950).

\bibitem{extensive2} A. Andronic, P. Braun-Munzinger, J. Stachel, Nucl. Phys. A{\bf 772}, 167 (2006).

\bibitem{Tsallis} C. Tsallis, J. Stat. Phys. {\bf 52}, 479 (1988).

\bibitem{non-extensive1}  T.S. Biro, G. Purcsel and K. Urmossy, Eur. Phys. J. A{\bf 40}, 325 (2009). 

\bibitem{non-extensive2} W.M. Alberico and A. Lavagno, Eur. Phys. J. A{\bf 40}, 313 (2009).

\bibitem{non-extensive3} T.S. Biro and G. Purcsel, Phys. Rev. Lett. {\bf 95}, 162302 (2005). 

\bibitem{urmossy} K. Urmossy, arXiv:1212.0260[hep-ph].

\bibitem{PPG099} A. Adare {\it et~al.} (PHENIX Collaboration), Phys.Rev. D{\bf 83}, 052004 (2011).

\bibitem{khandai_ijmpa} P.K. Khandai, P. Sett, P. Shukla and V. Singh,  Int. J. Mod. Phys. {\bf 28}, 1350066 (2013). 

\bibitem{tsallis_sps} M. Rybczynski and Z. Wlodarczyk, Eur. Phys. J. C{\bf 74}, 2785 (2014). 

\bibitem{tsallis_ch_hadrons1}  M.D. Azmi and J. Cleymans, J. Phys. G{\bf 41}, 065001 (2014). 

\bibitem{tsallis_ch_hadrons2} J. Cleymans {\it et~al.}, Phys. Lett. B{\bf 723}, 351 (2013). 

\bibitem{wong_wilk} C.Y. Wong and G. Wilk,  Phys. Rev. D{\bf 87}, 114007 (2013).

\bibitem{Hagedorn} R. Hagedorn, Riv. del Nuovo Cim. {\bf 6N 10}, 1 (1984).  

\bibitem{HAGEFACT} R. Blankenbecler and S. J. Brodsky, Phys. Rev. D{\bf 10}, 2973 (1974).

\bibitem{na61} N. Abgrall {\it et~al.}  Eur. Phys. C{\bf 74}, 2794 (2014).

\bibitem{PPG101}  A. Adare  {\it et~al.} (PHENIX Collaboration), Phys. Rev. C{\bf 83}, 064903 (2011).

\bibitem{cmspp_900_2.76_7} S. Chatrchyan  {\it et~al.} (CMS Collaboration), Eur. Phys. J. C{\bf 72}, 2164 (2012).

%\bibitem{pqcd_powerlaw} D.de Florian and R. Sassot, Phys. Rev. D{\bf 69}, 074028 (2004).

\bibitem{BG_statistics1} R. Hagedorn,  Nuovo Cim.  Suppl. {\bf 3}, 147 (1965).

\bibitem{BG_statistics2} D.B. Walton and J. Rafelski, Phys.  Rev. Lett. {\bf 84}, 31 (2000).

\bibitem{q_Tsallis} G. Wilk and Z. Wlodarczyk, Phys. Rev. Lett. {\bf 84}, 2770 (2000).

%\bibitem{Phenix} Adare A {\it et~al.} (PHENIX Collaboration),  Phys. Rev. C{\bf 83}, 064903 (2011).

\bibitem{Star} B.I. Abelev {\it et~al.} (STAR Collaboration), Phys. Rev. C{\bf 75}, 64901 (2007).

\bibitem{cleymans} J. Cleymans and D. Worku, Eur. Phys. J. A{\bf 48}, 160 (2012).

\bibitem{BlankenbeclerPRD12} R. Blankenbecler, S.J. Brodsky and J. Gunion, Phys. Rev. D{\bf12}, 3469 (1975) .

\bibitem{BrodskyPLB637} S.J. Brodsky, H.J. Pirner and J. Raufeisen, Phys. Lett. B{\bf 637}, 58 (2006).  


\bibitem{Tsallis-Pareto} T.S. Biro, K. Urmossy and G.G. Barnafoldi, J. Phys. G{\bf 35}, 044012 (2008).

\bibitem{Tsallis-Levy} K. Aamodt {\it et~al.} (ALICE Collaboration), Eur. Phys. J. C{\bf 71}, 1655 (2011).

%\bibitem{ppproton} B. I. Abelev {\it et~al.} (STAR Collaboration), Phys. Rev. C{\bf 75}, 64901 (2007). 
 
\bibitem{ppbaryon} J. Adams {\it et~al.}  (STAR Collaboration), Phys. Lett. B{\bf 637}, 161 (2006). %Phys. Rev. {\bf C 75}, 064901 (2007).
  
%\bibitem{pp62chargedpionkaonproton} A. Adare {\it et~al.}  (PHENIX Collaboration), Phys. Rev. C{\bf 83}, 064903 (2011).

%\bibitem{cms_had_corr} V. Khachatryan {\it et~al.}  (CMS Collaboration), JHEP {\bf 1009}, 091 (2010).  


%\bibitem{arxiv0812.1471} J. Cleymans, G. Hamar, P. Levai and S. Wheaton, J. Phys. G{\bf 36} 064018 (2009). 

%\bibitem{urmossy} K. Urmossy, arXiv:1212.0260[hep-ph].



\end{thebibliography}
\end{document}